# Electronic structure of (Ga,Mn)As revisited: an alternative view on the "Battle of the bands"

J. Kanski[*] and L. Ilver[†] Department of Applied Physics, Chalmers University of Technology, SE-412 96 Göteborg, Sweden

K. Karlsson[‡] Department of Engineering Sciences, University of Skövde, SE-541 28 Skövde, Sweden

M. Leandersson[§] MAX IV Laboratory, Lund University, SE-221 00 Lund, Sweden

I. Ulfat[¶] Department of Physics, University of Karachi, Karachi 75270, Pakistan

J. Sadowski[**] MAX IV Laboratory, Lund University, SE-221 00 Lund, Sweden and Institute of Physics, Polish Academy of Sciences, al. Lotnikow 32/46, PL-02-668 Warszawa, Poland

(Dated: October 31, 2014)

Abstract

New detailed angle-resolved photoemission data are presented, revealing the existence of an Mn-induced state that extends into the band gap of GaAs. In sharp contrast to recent reports we observe that the state is highly dispersive. Spin resolved photoemission shows that the band is spin polarized even at room temperature. The results are not consistent with any of the currently discussed band models for ferromagnetism.

e-mail addresses:
[*]janusz.kanski@chalmers.se, [†]lars.ilver@chalmers.se, [‡]krister.karlsson@his.se, [§]mats.leandersson@maxlab.lu.se, [¶]iulfat@uck.edu.pk, [**]janusz.sadowski@maxlab.lu.se

Introduction

Although more than 20 years have passed since the first synthetization of a III-V-based dilute magnetic semiconductor [1], implementation of these materials in everyday spin-based electronics is as elusive as ever because the ferromagnetic transition temperature is much too low. Rather remarkably, the physical origin of the ferromagnetic state is still debated, even for the prototype dilute magnetic semiconductor (Ga,Mn)As. While a wealth of experimental data suggests that the magnetic coupling is mediated by spin-polarized holes, the actual character of these holes has recently become an issue of debate. Two main scenarios are discussed: acceptor induced holes in the host valence band and holes in an impurity band. Experimental evidence for the existence of an impurity band based on optical properties has been presented [2], though later studies suggest that the data are consistent with the host valence band model [3]. Support for an impurity band scenario is also obtained from resonant tunneling experiments on quantum well structures [4]. Two other recent studies, one based on channeling in combination with magnetization, transport, and magneto-optical experiments, the other on hard X-ray photoemission have come to different conclusions: the first one supporting an impurity band model [5] in which the location of the Fermi level within the impurity band plays a crucial role in determining the Curie temperature ($T_C$), the second one emphasizing the coexistence of coupling mechanisms in the impurity band and host valence band models [6].

While photoemission is certainly the most direct probe of electronic states, its applicability is hampered by its intrinsic surface sensitivity: well-defined, atomically clean samples are required. This is not an issue in situations where the surface can be prepared by e.g. ion etching and annealing, but in the present case such treatment is prohibited because the material is metastable and undergoes phase separation at temperatures above 300 °C. An alternative is to protect the surface against contamination during transfer between the growth and analysis stages, e.g. by As capping. This is a very delicate method, since the overlayer must be sufficiently thick to serve its purpose, and at the same time possible to remove by heating. However, even if such heating is carried out under conditions avoiding phase separation, an additional complication specific for the present system is unavoidable: during post-growth annealing diffusing interstitial Mn in the (Ga,Mn)As sample will react with surface As to form MnAs overlayer/particles [7]. The only safe alternative avoiding these complications is *in situ* growth. For this reason we have connected a dedicated MBE growth system with the photoelectron spectrometer, allowing us to investigate as-grown samples transferred between the two units in ultrahigh vacuum. The fact that the results presented here have not been found in any previous study indicates that the sample handling is a decisive issue in this context.

Experiment

The experiments were carried out at MAX IV laboratory beamline I3, where a photoelectron spectrometer (Scienta R-4000) is directly connected to an MBE system (SVTA-N35). As already mentioned, this configuration allows us to transfer samples between the two units in ultrahigh vacuum. Photoelectrons are normally recorded with a microchannelplate/fluorescent screen assembly [8]. Alternatively, the electrons can be sent through an aperture next to the channelplate and to a mini-Mott spin detector. 50 nm thick (Ga,Mn)As layers with different Mn concentrations were grown on n-type GaAs(100) substrates (around 5x10 mm$^2$), which were glued with indium on Mo holders. The Mn concentrations were determined *in situ* by RHEED oscillations, which only probe Mn in substitutional sites [9], and were in some cases checked afterwards by means of secondary ion spectroscopy (SIMS). The RHEED oscillations were also used for defining a secondary *in situ* concentration scale based on Mn $2p_{3/2}$ absorption spectra. The XAS was recorded in total electron yield mode, which means a probing depth in the range of 5 nm [10]. A linear relationship was established [11] between the XAS amplitudes and the nominal concentrations over a wide range of concentrations. The concentrations quoted below are estimated to be accurate within 0.5%.

To facilitate as detailed comparison as possible between spectra from GaAs and (Ga,Mn)As, a mask was introduced in front of the sample after growth of the buffer LT GaAs layer, leaving only a part of the substrate exposed to the continued deposition. In this way a stripe of clean GaAs was left during subsequent growth of (Ga,Mn)As. As a beneficial side effect of the locally unbalanced beam fluxes, an approximately 1 mm wide metallic transition region was generated between the two areas. Thus, just by minute in-plane translations in front of the analyser we were able to record spectra from GaAs and (Ga,Mn)As and also determine the Fermi energy without changing any experimental parameters. Angularly dispersed photoemission spectra were recorded with an imaging detector system covering a range of ±15° around the surface normal. The overall experimental energy resolution was around 100 meV. All experiments were carried out with the sample at room temperature.

Results and discussion

All (Ga,Mn)As samples with Mn concentrations above 0.5% displayed (1x2) LEED patterns, and all GaAs surfaces showed c(4x4) reconstruction. The diffraction patterns from (Ga,Mn)As were characterized by somewhat higher background. In addition, the diffraction spots from (Ga,Mn)As were in general significantly broader than those from GaAs, indicating a smaller range of coherent scattering, i.e. lower degree of long range ordering. The (1x2) periodicity is believed to be a result of surface disorder: in a study of Mn incorporation in GaAs it was concluded that the lack of fractional order diffraction along the [110] direction is due to a mixture of (2xn)-type dimer units [12]. The notion of disorder was also invoked in an analysis of As core level photoemission and was directly supported by the observation of transition between the c(4x4) and (1x2) diffraction patterns at about 0.5% Mn concentration via broadening (streaking) of the quarter-order spots in the c(4x4) reconstruction pattern [13].

Figure 1 shows valence band angular distribution plots from (Ga,Mn)As with 1.2% Mn and the parallel GaAs surface, excited with 21 eV photons in p polarization. The angular scale can of course be easily transformed into an in-plane momentum scale. For example, at 16 eV kinetic energy and 10° emission angle the in-plane momentum corresponds to around 50% of the distance to the (1x1) surface Brillouin zone boundary. Energy distribution curves integrated over ±2° around the surface normal are also displayed on either side. Detailed comparison of spectra from the two systems requires that they be represented on a common energy scale. Due to the different doping situations it is of course not possible to use the Fermi level as a reference. The valence band maxima (VBM) are notoriously difficult to assess, not least because the spectra are dominated by surface state emission in this region. Also core levels are problematic: the As(3d) spectra contain several surface components and the Ga(3d) spectrum of (Ga,Mn)As is distorted by Mn-induced components [14]. In the present study we have used the $X_3$ "density of states" peaks (see e.g. ref. 15) as common reference levels. Since these structures are not dispersing, they can be located with good accuracy (within ± 30 meV) by integrating over the entire angular range of acceptance. Using this alignment, we find that the gross features in spectra from the two systems are indeed similar. The most apparent differences occur in the VBM region. This region is largely dominated by surface states [15], which complicates efforts to identify Mn-specific features. Still, less prominent, but for the present discussion most important differences can indeed be found in this region. It is clear, for example, that even though the shoulder at 16 eV kinetic energy (reflecting surface state emission in the spectrum of GaAs) is quenched in the spectrum of (Ga,Mn)As, this spectrum extends higher in energy than that of GaAs. While this tailing may be taken as an effect of disorder related broadening, it will be shown below that it actually reflects a rapidly dispersing Mn-induced band. Another feature that appears to be broadened is the bow-shaped structure around 13 eV kinetic energy, reflecting excitation the $\Delta_1$ band. However, also in this case the impression of broadening is misleading.

Some more details can be observed when the data are displayed in second derivative mode as in Figure 1b. Starting from the bottom it is seen that the aligned $X_3$ spectra, enlarged for clarity, do not show any broadening whatsoever. It can also be seen that the $X_5$ critical point emission (around 12 eV) coincides in energy as a result of the alignment of the $X_3$ structures. The spectrum of the $\Delta_1$ band (integrated over ±2° around the surface normal) appears indeed broader in the spectrum from (Ga,Mn)As. More detailed inspection reveals that the structure is not simply broadened, but is clearly asymmetric. Numerical simulations of second derivative spectra show that the observed shape is consistent with the presence of a double peak. It can be noted that the low-energy shoulder coincides with the corresponding peak in the spectrum from GaAs. Interestingly, a splitting of the spectral function corresponding to the $\Delta_1$ band has been found in recent *ab initio* calculations, reminiscent of the present observation [16]. In this context it is also interesting to note that in an early photoemission study [17] a shift between the $\Delta_1$ bands in GaAs and (Ga,Mn)As was reported, though in the opposite direction relative the apparent shift found here. The discrepancy can probably be ascribed to different spectral alignments. Nevertheless, it is worth to emphasize that when a dispersing band like the $\Delta_1$ in GaAs is shifted in energy, the underlying perturbation is not local (i.e. confined to individual Mn atoms and their nearest four As neighbours). In other words, with impurity concentrations in the range of 1% the electronic structure is globally modified and thus cannot be described trivially in terms of doping.

The second derivative display reveals more extensive Mn-related changes of the VBM region than indicated above. The quenching of the GaAs(001)c(4x4) surface state band is obvious, but some additional, more subtle modifications can be discerned in the vicinity of normal emission. Specifically we notice a triangular, somewhat less bright feature with its top around 15.7 eV kinetic energy within the "eye-shaped" contour in the GaAs figure. For (Ga,Mn)As a corresponding feature is seen to extend through the border of the eye region and extend up to the Fermi level.

Before entering a more detailed discussion of the VBM data, we shall summarize briefly the results of earlier photoemission experiments focusing on this all-important energy region. In the aforementioned study [17] a set of normal emission spectra was presented and a non-dispersive Mn-induced state was found just below VBM. A more recent study, employing high photon energies (3.2 keV), reported an Mn induced structure above the VBM of GaAs [6]. No in-plane dispersion was observed, probably because of insufficient angular resolution. (A resolution better than 0.1° would be needed to resolve features corresponding to those displayed below). Two other recent studies should be mentioned in this context, both employing resonant photoemission [18, 19]. Mn-induced states were found only below VBM in both cases. It is noted, however, that resonantly enhanced photoemission projects out local states with a given symmetry (l=2 in this case). The absence of resonant enhancement above VBM shows that the dispersive band above VBM discussed below is not derived from Mn 3d states.

In Figure 2 we show spectra of the VBM region from the GaAs and (Ga,Mn)As samples. All data in this figure are displayed on the recorded kinetic energy scale, i.e. they are not aligned as in Figure 1. In agreement with many previous studies of GaAs, the Fermi level is pinned at the surface near midgap and the gap region is completely free from photoelectrons even in the plot with minimal threshold level (Figure 2b). As anticipated from the survey data, the emission from the (Ga,Mn)As spectrum extends towards higher energies, see Figure 2c. Since the emission falls rapidly in intensity, it is not possible to display the details without totally overexposing the image. To get around this complication we have composed an image from slices, each adjusted arbitrarily with respect to threshold and saturation levels, see Figure 2d). With this it becomes clear that the spectral tailing towards high energy noticed in Figure 1 actually reflects a well-defined energy band that reaches the Fermi level. So, unlike all previous studies we are able to directly detect delocalized electron states that are specific for (Ga,Mn)As. In Figure 2d we have also indicated the VBM position in GaAs (dotted line), taking into account the different pinning situations in the two materials. Its location in energy was estimated using literature data [20, 21], according to which the energy separation between the $X_3$ point and VBM is in the range 6.7 – 6.9 eV. The dotted line marks the highest possible position based on these data, so it can be safely concluded that the narrow band extends into the band gap region of GaAs.

Having established the existence of an Mn-induced band, it is important to further examine its properties. Of particular concern is the possibility that our observation represents a surface state. This issue can be addressed in several ways. A surface state is of course two-dimensional, a property that can be easily checked by comparing spectra recorded with different photon energies while keeping a fixed in-plane momentum. Our data show indeed very little, if any dependence of momentum along surface normal, which is the signature of a surface state. While this is normally a reliable identification, the present situation turns out to be more complicated as other observations contradict such assignment. First, we see in Figure 2 that the band is not confined to the band gap region, but can be followed at least 1 eV below VBM (not shown here). Second, referring to the surface geometry, we find that the in-plane dispersion is completely isotropic - no asymmetry can be discerned that could be connected with the (1x2) surface reconstruction. Third, a well defined and rapidly dispersing surface state band would indicate a very well ordered surface with long-range coherence. As already mentioned, however, the (1x2) reconstructed surface is characterised by disorder. Furthermore, the (1x2) electron diffraction pattern is in general relatively diffuse and no correlation has been observed between the quality of the diffraction pattern and appearance of the Mn-induced band. In fact the band has been found quite stable against contamination, such that it is clearly observed even when the most prominent bulk derived features in the energy distribution curves are severely attenuated. All this leads us to conclude that the conspicuous energy band found in the data from (Ga,Mn)As is not a regular surface state band. As will be argued below, the observations can be explained by the presence of a surface layer of (Ga,Mn)As with qualitatively different properties than the underlying bulk.

It is known [22] that the ferromagnetic state in (Ga,Mn)As appears only at Mn concentrations above approximately 1 %. Our 1.2 % sample with a $T_C$ of around 10K is thus a borderline case. With this in mind it is noted that the Mn induced band reaches the Fermi level and is obviously a candidate for providing the delocalized holes needed for ferromagnetic coupling. The question is then whether the band exhibits a concentration dependence that is consistent with the known magnetic properties. In Figure 3 we show the VBM intensity plots for three samples with different Mn nominal concentrations, 0.5%, 1.2% and 5.0 %. The 5 % sample showed a $T_C$ of 55 K, typical for our as-grown samples in this concentration range, while for the 0.5 % sample remained paramagnetic to below 5K. Some important observations can be made. First of all, we note that the Mn-induced state appears in all cases. In this context it must be stressed that the accuracy of the quoted Mn concentrations is only within 0.5 %, and that we have no detailed information regarding the point at which the new band appears. Nevertheless it can be mentioned that in data from a sample with a nominal concentration of 0.25 % (not shown here) a weak signature of the band can be discerned as a sharpened shape of the quite flat VBM emission contour of pure GaAs. This limited accuracy has no impact on the observed qualitative trend: for concentrations below 1% the Mn-induced band does not reach the Fermi level, and can therefore not host any delocalized hole states. In the data from the 5% sample the two branches of the band appear more separated, which we interpret as an effect of a shift of the band towards high energy. With increasing Mn density the band can host an increasing density of holes, which clearly parallels the concentration dependence of $T_C$.

The actual origin of the Mn-induced band remains to be clarified via detailed theoretical analysis. The strong dispersion, independent of Mn concentration, shows that it cannot occur as a consequence of overlapping impurity states, as proposed in the impurity band model. In connection with Figure 1 we commented an Mn-induced change within the eye-shaped profile, suggesting that the emission reaching the Fermi level actually developes from the GaAs band that has a maximum energy e few tenths of an eV below VBM. A tentative interpretation would thus be that the band appearing in (Ga,Mn)As has its origin in the band structure of GaAs, specifically in the spin-orbit split band that has its maximum energy around 0.4 eV below VBM.

To our knowledge there are two theoretical papers in which some indications of "anomalous" band features in the VBM region are indicated [16, 23]. In both cases an excursion of a host-derived majority spin band is predicted above VBM. In ref. 16 an explanation is offered in terms alloy disorder effect on the host band structure around the centre of the Brillouin zone. At this point it must be stressed that both calculations treat (Ga,Mn)As in its ferromagnetic state, while the experimental data discussed here were all recorded at room temperature. Knowing that $T_C$ is always far below room temperature for all samples, we are facing a very surprising situation. How can we understand the correspondence between experimental data on paramagnetic (Ga,Mn)As and theoretical results on the ferromagnetic system? A simple explanation might be that the electronic structures of (Ga,Mn)As in para- and ferromagnetic states are qualitatively similar. We are not aware of electronic structure calculations of paramagnetic (Ga,Mn)As and are thus not able to explore this possibility further, although intuitively it seems unlikely. There is another way in which the apparent controversy can be resolved. Since magnetization measurements (using e.g. squid instrumentation) sense the magnetic moment of the whole sample, while photoemission is intrinsically a surface sensitive probe, an obvious solution is that the surface of as-grown (Ga,Mn)As is ferromagnetic, but the volume of this phase is too small to be detected in a squid experiment. To examine the viability of this hypothesis, we need to know the sensitivity of the squid experiment in terms of number of (Ga,Mn)As layers. The magnetic moment per Mn atom can be estimated from literature data. In a typical magnetization study of post-growth annealed (Ga,Mn)As with 7% Mn [24], the magnetization is found to be in the range of 10 emu/cm$^3$. In this case the effective magnetic moment per Mn atom is extracted to be around 0.7 $\mu_B$. For a sample area of 5x5 mm$^2$ and x % Mn we then find that the magnetic moment per atomic layer is 10$^{-6}$ x emu. Taking into account that the effective moment per Mn atom in our as-grown samples is likely to be lower than that used in this calculation (because the remaining interstitial Mn reduces the magnetization via antiferromagnetic ordering), we can conclude that a typical sensitivity of 10$^{-7}$ – 10$^{-8}$ emu may indeed be insufficient to detect a couple of ferromagnetic monolayers.

Since our photoelectron spectrometer is equipped with a Mott spin polarimeter, the above ideas can be tested experimentally. According to calculations [16, 23] the band extending above VBM should contain only majority spin electrons, so we focus on this energy region. A problem in these measurements is of course the very low spectral intensity and the relatively low efficiency of the Mott detector (the Sherman function is estimated to be only around 15%). Nevertheless, reproducible data could be recorded, as shown in Figure 4 for a sample with 5% Mn. With the expected easy axis along the [-110] direction [25], the sample was oriented such that this direction (together with the surface normal) was in the scattering surface plane of the Mott detector. The degree of spin polarization was obtained by combining the scattered intensities in pairwise detectors following standard procedures [26]. Thus we find no sign of spin polarization along the surface normal, see Figure 4a. However, the corresponding plot for the in-plane polarization shows clear deviation from the zero line just in the region where the Mn-induced band appears above the VBM of GaAs, see Figure 4b. The signal is admittedly noisy, nevertheless it is reproducible and reliable. We can thus conclude that the energy band discussed above is indeed spin polarized. With this in mind, we can understand why this band appears significantly more distinct than other states: it belongs to the majority spin population, so in the spin polarized system the phase space for inelastic scattering is significantly smaller than that for electrons in minority spin state.

Conclusion

We conclude that the surface of (Ga,Mn)As is ferromagnetic even at room temperature and, as indicated by the two-dimensional characteristics of the Mn-induced energy band, the ferromagnetic phase is confined to a very thin surface region. The conclusion is consistent with the empirical similarities between the experimental data and theoretical band structure calculations on ferromagnetic (Ga,Mn)As. It deserves to be pointed out that the occurrence of a ferromagnetic surface may explain one of the open issues regarding the properties of (Ga,Mn)As, namely the unidirectional magnetic anisotropy (i.e. magnetic inequivalence of [110] and [-110] crystallographic directions) [25].

Our experimental results on (Ga,Mn)As differ radically from the prevailing view of this model semiconductor system. The Mn-induced modifications found here are of a different nature than discussed in the host- or impurity band models and "Battle of the bands" [27] appears to have more than the two contestants. The presence of a ferromagnetic surface layer even at room temperature is totally unexpected. It is indeed surprising that after nearly two decades of intensive studies by several groups such information has been overlooked.

References

[1] H. Munekata, H. Ohno, S. von Molnar, Armin Segmüller, L.L. Chang, and L. Esaki, Phys. Rev. Lett. **63**, 1849 (1989)

[2] K. S. Burch, D.B. Shrekenhamer, E.J. Singley, J. Stephens, B.L. Sheu, R.K. Kawakami, P. Shiffer, N. Samarth, D.D. Awschalom, and D.N. Basov, Phys. Rev. Letters **97**, 087208 (2006)

[3] T. Jungwirth, P. Horodyska, N. Tesarova, P. Nemec, J. Subrt, P. Maly, P. Kuzel, C. Kadlec, J. Masek, I. Nemec, M. Orlita, V. Novak, K. Olejnik, Z. Soban, P. Vasek, P. Svoboda, and Jairo Sinova, Phys. Rev. Letters **105**, 227201 (2010)

[4] S. Ohya, K. Tanaka, and M. Tanaka, Nature Physics **7**, 342 (2011)

[5] M. Dobrowolska, K. Tivakornsasithorn, X. Liu, J. K. Furdyna, M. Berci, K. M. Yu, and W. Walukiewicz, Nature Materials **11**, 444 (2012)

[6] A. X. Gray, J. Minár, S. Ueda, P. R. Stone, Y. Yamashita, J. Fujii, J. Braun, L. Plucinski, C. M. Schneider, G. Panaccione, H. Ebert, O. D. Dubon, K. Kobayashi, and C. S. Fadley, Nature Mater. **11**, 957 (2012)

[7] J. Sadowski, M.Adell, J. Kanski, L. Ilver, E. Janik, E. Lusakowska, J. Z. Domagala, S. Kret, P. Dluzewski, R. Brucas, M. Hanson, "Self-organized MnAs quantum dotsformed during annealing of GaMnAs under arsenic capping", Appl. Phys. Lett. **87**, 263114 (2005)

[8] M. H. Berntsen, P. Palmgren, M. Leandersson, A. Hahlin, J. Åhlund, B. Wannberg, M. Månsson and O. Tjernberg, Rev. Sci. Instrum. **81**, 035104 (2010)J.

[9] Sadowski, J.Z. Domagala, J. Bak-Misiuk, S. Kolesnik, M. Sawicki, K. Swiatek, J. Kanski, L. Ilver, and V. Ström, J. Vac. Sci. Technol. B **18**, 1967 (2000)

[10] B. H. Frazer, B. Gilbert, B. R. Sonderegger, and G. De Stasio, Surf. Sci. **537**, 161 (2003).

[11] I. Ulfat, J. Kanski, L. Ilver, J. Sadowski, K. Karlsson, A. Ernst, and L. Sandratskii, Phys. Rev. B **89**, 04512 (2014)

[12] I. Ulfat, J. Adell, J. Sadowski, L. Ilver, and J. Kanski, Surface Sci. **604**, 125 (2010)

[13] A. Othake, A. Hagiwara, J. Nakamura, Phys. Rev. B **87**, 165301 (2013)

[14] J. Kanski, I. Ulfat, L. Ilver, M. Leandersson, J. Sadowski, K. Karlsson, and P. Pal, J. Phys. Condens. Matter **24**, 435802 (2012)

[15] P.K. Larsen, J.H. Neave, J.F. van der Veen, P.J. Dobson, and B.A. Joyce, Phys. Rev. B**27**, 4966 (1983)

[16] K. Sato, L. Bergqvist, J. Kudrnovsky, P.H. Dederichs, O. Eriksson, I. Turek, B. Sanyal, G. Bouzerar, H. Katayama-Yoshida, V.A. Dinh, T. Fukushima, H. Kizaki, and R. Zeller, Rev. Mod. Phys. **82**, 1633 (2010)

[17] J. Okabayashi, A. Kimura, O. Rader, T. Mizokawa, A. Fujimori, T. Hayashi, and M. Tanaka, Phys. Rev. B **66**, 125304 (2001)

[18] M. Kobayashi, I. Muneta, Y. Takeda, Y. Harada, A. Fujimori, J. Krempaský, T. Schmitt, S. Ohya, M. Tanaka, M. Oshima, and V.N. Strocov, Phys. Rev. B **89**, 205204 (2014)

[19] I. Di Marco, P. Thunström, M. I. Katsnelson, J. Sadowski, K. Karlsson, S. Lebègue, J. Kanski, and O. Eriksson, Nature Commun. **4**, 2645 (2013)

[20] T.-C. Chiang, R. Ludeke, A. Aono, G. Landgren, F.J. Himpsel, and D.E. Eastman, Phys. Rev. B**27**, 4770 (1983)

[21] G.P. Williams, F. Cerrina, G.J. Lapeyre, J.R. Anderson, R.J. Smith, and J. Hermanson, Phys. Rev. B**34**, 5548 (1986)

[22] see e.g. T. Dietl and H. Ohno, Materials Today **9**, 18 (2006)

[23] A. Ernst, L.M. Sandratskii, M. Bouhassoune, J. Henk, and M. Lüders, Phys. Rev. Lett. **95**, 237207 (2005)

[24] P. Nemec, V. Novák, N. Tesarová, E. Rozkotová, H. Reichlová, D. Butkovicová, F. Trojánek, K. Olejník, P. Malý, R.P. Campion, B.L. Gallagher, Jairo Sinova, and T. Jungwirth, Nature Commun. **4**, 1422 (2013)

[25] J. Zemen, J. Kucera, K. Olejník, and T. Jungwirth, Phys. Rev. B **80**, 155203 (2009)

[26] see e.g. J. Stöhr and H.C. Siegmann, 2006, *Magnetism From Fundamentals to Nanoscale Dynamics* (Springer Verlag Berlin Heidelberg)

[27] N. Samarth, Nature Materials **11**, 360 (2012)

Figure captions

Figure 1 a) Valence band intensity plot from GaAs(100)-c(4x4) excited with p-polarized 21 eV photons and recorded along the [-110] azimuth. b) Corresponding intensity plot from (Ga,Mn)As(100)-(1x2) with 1.2 % Mn. The curves on either side show the energy distribution curves obtained by integrating the respective images over ± 2° around the surface normal. The spectra are shifted in energy to compensate for the different doping situations in the two cases. The Fermi level is indicated in each case with a white line. c) and d) Second derivative presentation of the same data as in a) and b).

Figure 2 Close-up view of the VBM region from a) GaAs and c) (Ga,Mn)As. b) An amplified image of the GaAs data and d) an image of the (Ga,Mn)As data composed of slices with gradually increasing amplification towards the Fermi level. The dashed line indicates the Fermi level and the dotted line in d) represents the valence band maximum of GaAs as described in the text.

Figure 3 VBM data from (Ga,Mn)As samples with a) 0.5 % (excited with 24 eV photons), b) 1.2 % and c) 5 % Mn (both excited with 25 eV photons). The dashed line indicates the Fermi level.

Figure 4 Spin polarization along surface normal (a) and in-plane (b) along [-110] azimuth for a sample with 5 % Mn. The data were recorded at room temperature.

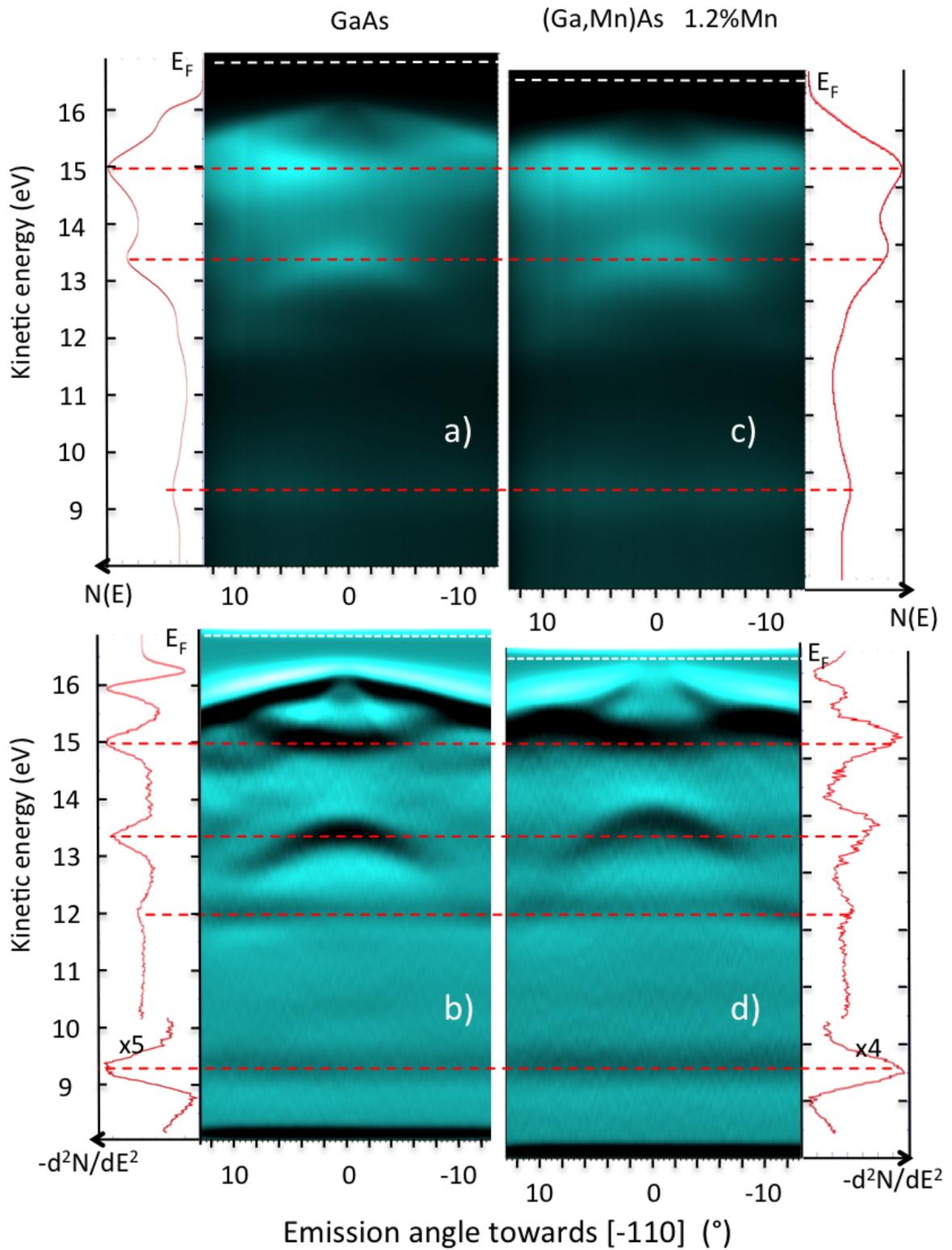

Figure 1

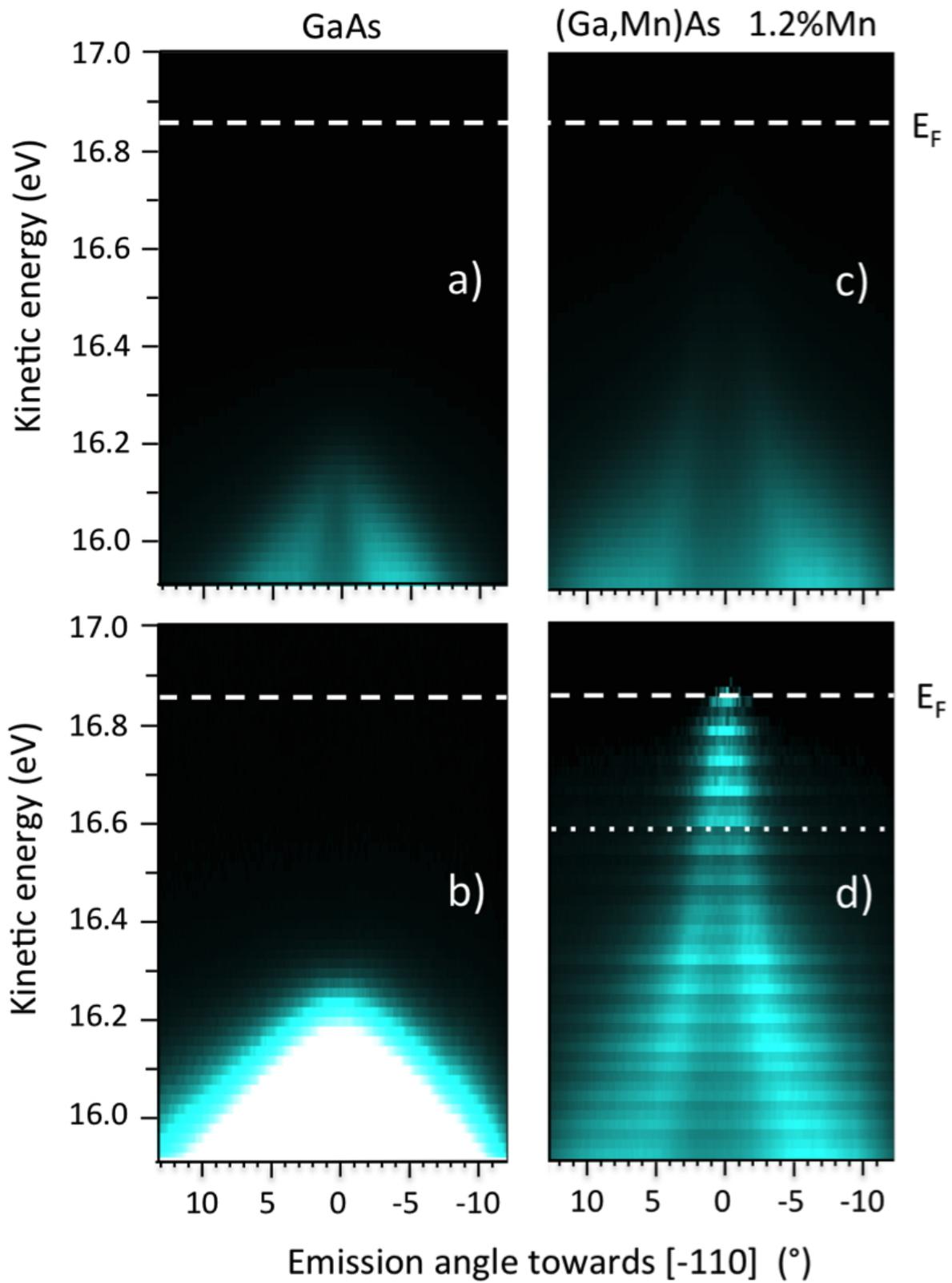

Figure 2

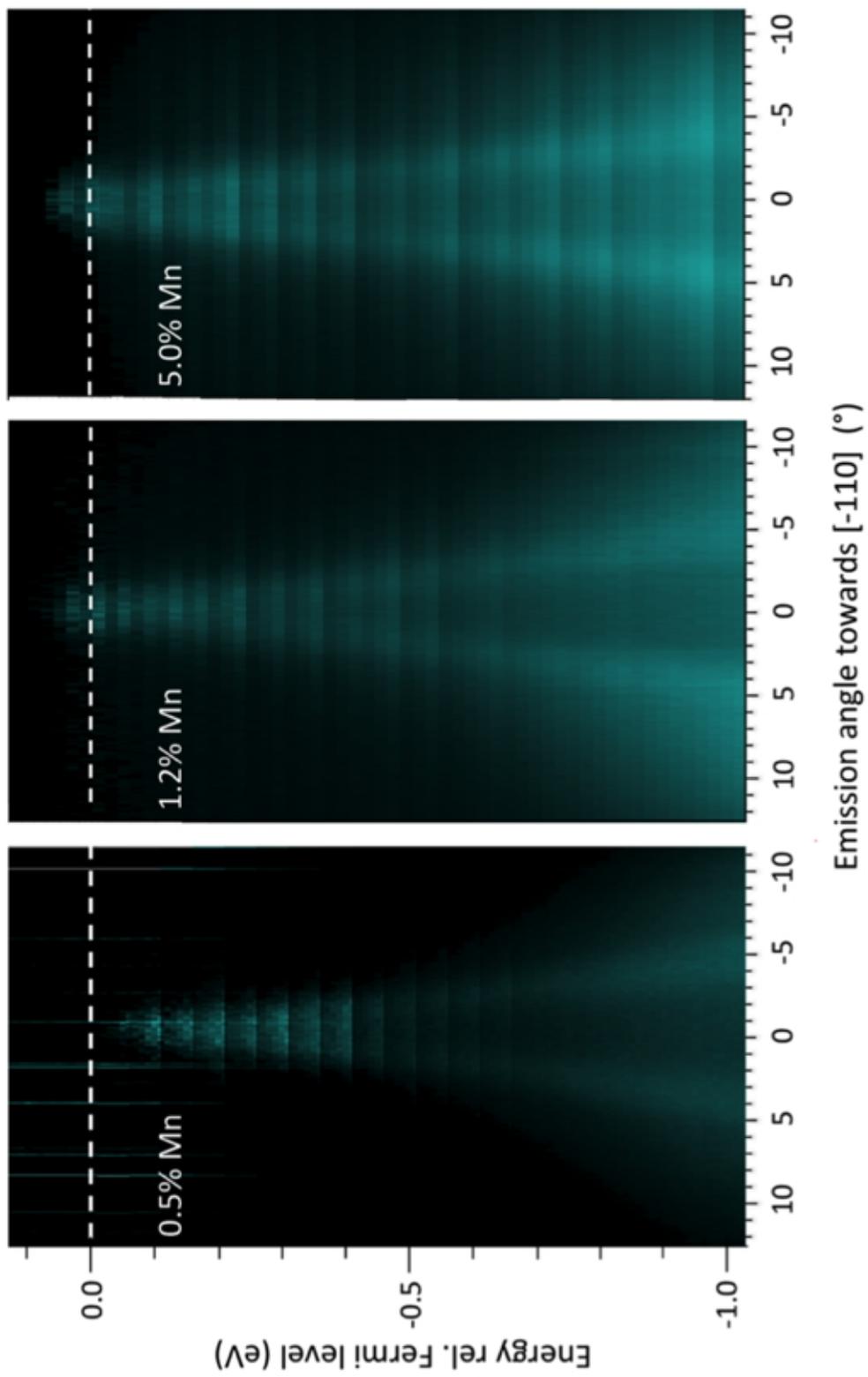

Figure 3

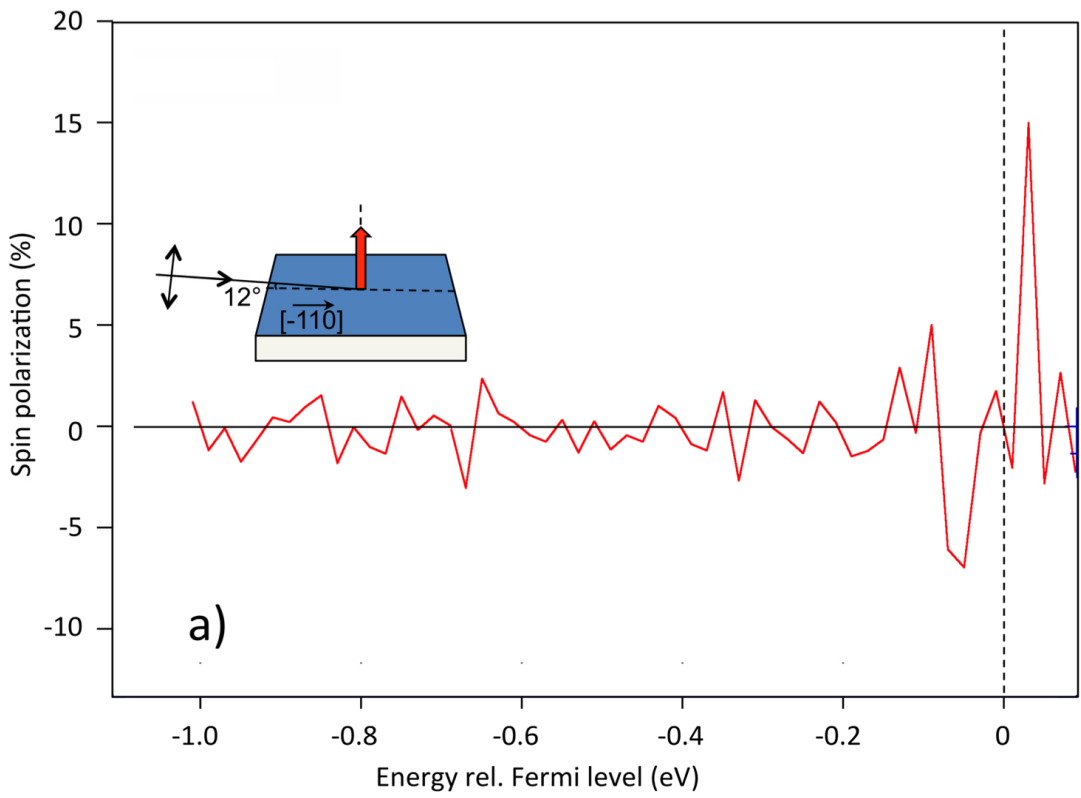

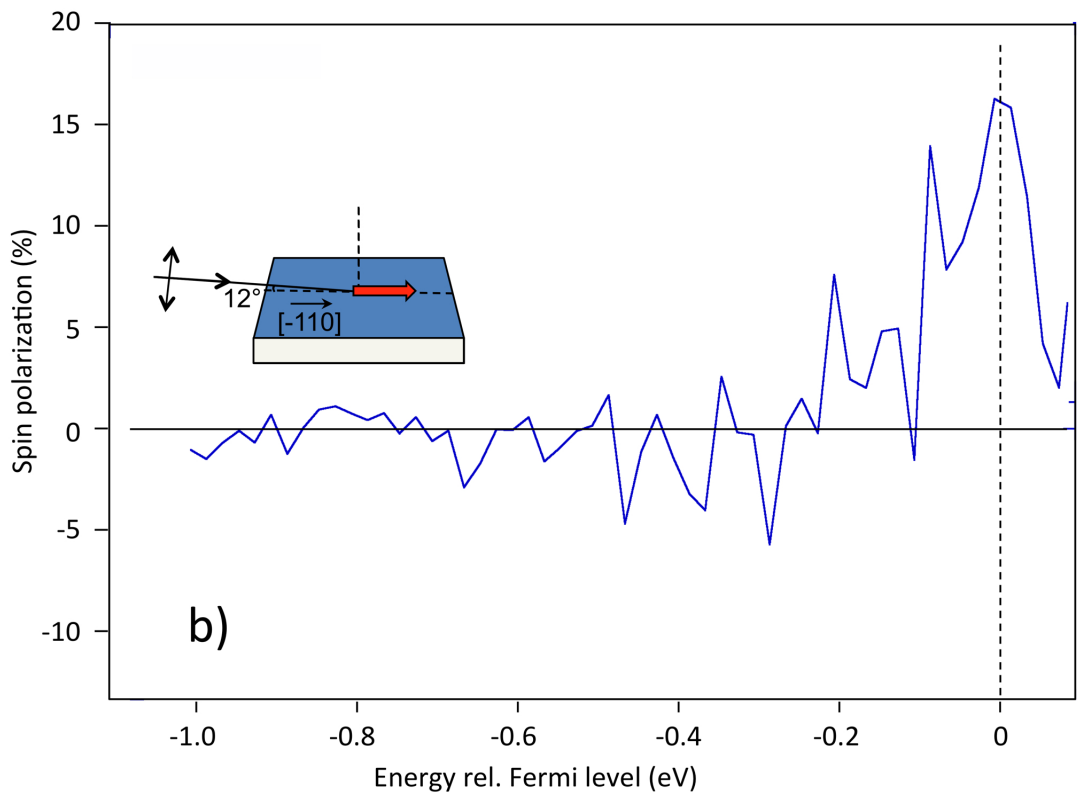

Figure 4